\documentclass[final,5p]{elsarticle}
\biboptions{sort&compress}

\usepackage[dvipsnames]{xcolor}

\usepackage{amssymb}
\usepackage{bbm}
\usepackage{bm}
\usepackage{amssymb,graphicx}
\usepackage[intlimits]{amsmath}
\usepackage[small]{subfigure}
\usepackage{color}
\usepackage{array}
\usepackage{multirow}
\usepackage{dsfont}
\usepackage[colorlinks=true]{hyperref}
\hypersetup{
 linkcolor=blue,citecolor=blue,urlcolor=black}
 
\allowdisplaybreaks

\newcommand{\su}{\mathfrak{su}}
\newcommand\eq[1]{eq.~(\ref{eq:#1})}

\DeclareMathOperator{\tr}{tr} 

\newcommand{\iu}{\text{i}}

\newcommand{\CB}{\mathcal{B}}
\newcommand{\CF}{\mathcal{F}}
\newcommand{\CL}{\mathcal{L}}
\newcommand{\CN}{\mathcal{N}}
\newcommand{\CO}{\mathcal{O}}

\newcommand{\Fg}{\mathfrak{g}}
\newcommand{\Fh}{\mathfrak{h}}
\newcommand{\Fgl}{\mathfrak{gl}}
\newcommand{\Fo}{\mathfrak{o}}
\newcommand{\Fso}{\mathfrak{so}}

\newcommand{\Fsu}{\mathfrak{su}}
\newcommand{\Fu}{\mathfrak{u}}

\journal{Physics Letters B}

\begin{document}

\begin{frontmatter}

\title{Integrable Corners in the Space of Gukov-Witten Surface Defects}

\author[cph]{Adam Chalabi}
\author[cph]{Charlotte Kristjansen}
\author[cph]{Chenliang Su}

\affiliation[cph]{organization={Niels Bohr Institute, Copenhagen University},
            addressline={Blegdamsvej 17}, 
            city={Copenhagen},
            postcode={2100}, 
            country={Denmark}}
\ead{adam.chalabi@nbi.ku.dk, 
kristjan@nbi.dk,
chenliang.su@nbi.ku.dk
}

\begin{abstract}
We investigate integrability properties of Gukov-Witten 1/2-BPS surface defects in $SU(N)$ $\mathcal{N}=4$ super-Yang-Mills (SYM) theory in the large-$N$ limit.
We demonstrate that ordinary Gukov-Witten defects, which depend on a set of continuous parameters, are not integrable except for special sub-sectors.
In contrast to these, we show that rigid Gukov-Witten defects, which depend on a discrete parameter but not on continuous ones, appear integrable in a corner of the discrete parameter space.
Whenever we find an integrable sector, we derive a closed-form factorised expression for the leading-order one-point function of unprotected operators built out of the adjoint scalars of $\mathcal{N}=4$ SYM theory. 
Our results raise the possibility of finding an all-loop formula for one-point functions of unprotected operators in the presence of a rigid Gukov-Witten defect at the corner in parameter space.
\end{abstract}


\end{frontmatter}

\section{Introduction\label{sec:intro}}

Extended operators, or defects, are non-local probes of quantum field theory (QFT).
In 4d gauge theory, defects play an essential role in characterising the theory and its vacua.
E.g.\ it is well-known that the vacuum expectation values (VEVs) of line operators on closed loops classify vacua as Coulomb, Higgs, or confining. 
In this work we focus on a type of defects that has been explored to a much lesser extent: surface defects in 4d gauge theory.

Surface defects provide order parameters that allow for discerning phases of gauge theories that line operators cannot distinguish~\cite{Gukov:2013zka}.
They can have various constructions and wildly different properties.
In pure abelian gauge theory, the simplest surface defects are engineered by prescribing a singularity in the $U(1)$ gauge field, or by turning on a 2d theta angle.
In the present manuscript, we study versions of these defects in non-abelian and maximally supersymmetric (SUSY) gauge theory, namely Gukov-Witten defects in $SU(N)$ $\CN=4$ super-Yang-Mills (SYM) theory~\cite{Gukov:2006jk,Gukov:2008sn}.
These defects preserve exactly half of the SUSY and superconformal generators of $\CN=4$ SYM theory.

For defects in $\CN=4$ SYM theory, a large number of non-perturbative tools are available.
They include large-$N$ integrability, SUSY localisation, and the (super)conformal bootstrap.
In particular, there has been a rich interplay between integrability and localisation, with developments using the former method often inspiring results in the same system using the latter, and vice versa.
Most efforts so far have focussed on line defects and boundaries, or interfaces, in $\CN=4$ SYM theory.
E.g.\ localisation results for correlation functions of protected local operators with an 't Hooft loop insertion~\cite{Okuyama:2006jc,Kristjansen:2023ysz} sparked the investigation of unprotected operators in the same system and its integrability properties~\cite{Kristjansen:2023ysz,Kristjansen:2024dnm}. 
Conversely, the application of integrability methods to the study of correlation functions in the presence of domain wall defects in ${\cal N}=4$ SYM theory~\cite{deLeeuw:2015hxa,deLeeuw:2016ofj,Buhl-Mortensen:2017ind} inspired the development of localisation methods for that system~\cite{Wang:2020seq, Komatsu:2020sup}.

In comparison, surface defects in $\CN=4$ SYM theory have received less attention. 
Early investigations of Gukov-Witten defects were primarily holographic~\cite{Gomis:2007fi,Drukker:2008wr}.
Although SUSY localisation methods were applied in an indirect way to this system in ref.~\cite{Chalabi:2020iie}, it was only recently that ref.~\cite{Choi:2024ktc} proposed an inherent SUSY localisation prescription.
The natural question arises whether integrability has a role to play for those systems as well. 

We address this question in the present paper. 
It has previously been suggested that some surface defects may be integrable~\cite{deLeeuw:2024qki}.
In this work, we will show that generic Gukov-Witten defects are not integrable but they are likely to become integrable at a corner of parameter space. 

Our paper is organised as follows. 
In section~\ref{Surfaces} we introduce Gukov-Witten surface defects.
We distinguish two types of surface defects, ordinary and rigid Gukov-Witten defects, with the latter arising at a special point in the parameter space of the former. 
In section~\ref{sec:int_tools} we briefly review the integrability tools that we will apply in our analysis of these defects. 
Sections~\ref{ordinary} and~\ref{rigid} contain our results for the ordinary and the rigid case, respectively. 
We conclude with a discussion of further perspectives in section~\ref{Outlook}.

\emph{Note added:} 
During the final stages of the present work we became aware of ref.~\cite{Holguin:2025bfe}, which also discusses the integrability properties of Gukov-Witten defects. 
The results presented here go beyond those of ref.~\cite{Holguin:2025bfe}.

\section{Surface defects in $\CN=4$ SYM theory \label{Surfaces}}

Throughout this paper we consider $SU(N)$ $\CN=4$ SYM theory on $\mathds{R}^4$ with $N$ taken to be large.
We will assume that the surface defect is supported on $\mathds{R}^2\subset\mathds{R}^4$.
The radial and angular coordinates in the directions transverse to the defect are denoted by $r$ and $\psi$, respectively.

We distinguish two types of Gukov-Witten defects, which we call ordinary and rigid, depending on whether they admit continuous parameters (ordinary) or not (rigid).

\subsection{Ordinary Gukov-Witten defects}

Ordinary Gukov-Witten defects in $SU(N)$ $\CN=4$ SYM theory are characterised by a scale-invariant singularity in some of the ambient fields at the support of the defect~\cite{Gukov:2006jk}.
Given a partition of $N=\sum_{m=1}^M N_m$, the prescribed singularity can depend on $4M$ continuous real parameters, $(\alpha_m,\beta_m,\gamma_m,\eta_m)$, where $m=1,\ldots, M$, which break the gauge group $SU(N)\to \mathds{L}=S(\prod_{m=1}^MU(N_m))$.
These singularities are consistent with preserving half of the superconformal charges of the ambient theory.
As shown explicitly in ref.~\cite{Gomis:2007fi}, there are corresponding superconformal background field configurations.
Concretely, they are characterised by a non-trivial background for the angular component of the gauge-field and for a single complex scalar field $\Phi=(\phi_1+\iu\phi_2)/\sqrt{2}$. 
For the scalar we have
\begin{equation}\label{ZAbelian}
\Phi = \frac{1}{\sqrt{2}z} \text{diag}((\beta_1+\iu \gamma_1)  \mathds{1}_{N_1}, \ldots, (\beta_M+\iu \gamma_M)  \mathds{1}_{N_M})\,,
\end{equation}
where $z=r e^{\iu\psi}$ and $\mathds{1}_{N_m}$ is the $N_m\times N_m$ identity matrix.
The classical background of the gauge field takes the form of a non-abelian vortex configuration
\begin{equation}
A= \text{diag}(\alpha_1 \mathds{1}_{N_1}, \alpha_2 {\mathds 1}_{N_2},\ldots, \alpha_M  {\mathds 1}_{N_M} )d \psi
\end{equation}
with the $\alpha_m$ real and periodic.
The remaining $M$ parameters $\eta_m$ are 2d theta-angles on the defect, which will play no role here.
Such a defect preserves a 
$(PSU(1,1|2)\times PSU(1,1|2))\rtimes SO(2)$ subgroup of the $ PSU(2,2|4)$ superconformal group of $\CN=4$ SYM theory.

The defect has a dual description as a probe D3 brane with geometry $AdS_3\times S^1\subset AdS_5\times S^5$~\cite{Drukker:2008wr,Constable:2002xt} where the intersection of $AdS_3$ with the boundary of $AdS_5$ generates the surface defect. 
Furthermore, the $AdS_3$ has an arbitrary inclination angle 
with respect to the $AdS_5$ boundary.
The probe has the additional feature that the $S^1$ is embedded non-trivially in $AdS_5\times S^5$, giving the probe a cone-like shape whose base is the $S^1$.
This is in contrast to the more studied probe D3-D5 and D3-D7 configurations with geometry $AdS_4 \times S^2$ and $AdS_4 \times S^4$, leading to co-dimension one defects, where only a dependence on the $AdS_5$ radius appears in the embedding. 
Dekel and Oz have provided a list of integrable D-brane configurations in $AdS_5\times S^5$~\cite{Dekel:2011ja}. 
For these configurations the probe brane always has an inclination of $\pi/2$ with respect to the $AdS_5$-boundary and the world volume coordinates have no dependence on any $AdS_5$-angle.
The argument of Dekel and Oz has been extended to D-branes with a general inclination angle in some cases~\cite{Linardopoulos:2021rfq}. 
However, ordinary Gukov-Witten defects still do not fit in the Dekel-Oz classification due to the non-trivial embedding of the $S^1$.

\subsection{Rigid Gukov-Witten defects}
Rigid Gukov-Witten defects in $SU(N)$  ${\cal N} = 4$ SYM theory can be obtained by taking a
limit of the parameters $\alpha, \beta, \gamma \to 0$, such that the leading singularity in the scalar fields becomes weaker than $1/r$~\cite{Gukov:2006jk, Gukov:2008sn}. 
Near the $r=0$ singularity the fields behave as\footnote{
Our conventions differ from the ones of ref.~\cite{Gukov:2006jk}, which are gauge-equivalent.
}
\begin{equation}
\label{eq:rigidGW}
A=\frac{t_3}{\log \frac{r}{r_0}} d\psi\,, \hspace{0.5cm} \Phi=\frac{t_1 +\iu t_2}{\sqrt{2}z\log\frac{r}{r_0}} \,,
\end{equation}
where  $t_1,t_2,t_3$ constitute a $k$-dimensional irreducible representation of $\su(2)$.
The scale $r_0$ should be interpreted as a dynamical field and should be integrated over in the path integral, restoring the system's conformal invariance~\cite{Gukov:2008sn}.
The resulting defect preserves a larger symmetry algebra than the ordinary Gukov-Witten defect.
Any transverse $SO(2)$ rotation and any $SO(2)$ R-symmetry transformation acting on $(\phi_1,\phi_2)$ can be undone by a constant $SU(2)\subset SU(N)$ gauge transformation.
Therefore, both transverse spacetime and R-symmetry $SO(2)$ factors are separately conserved, and so the symmetry is enlarged to $SU(1,1|2)\times SU(1,1|2)$.
In the dual probe brane picture the base of the cone shrinks to zero size, removing the dependence of the brane configuration on the $AdS_5$ angular coordinates and possibly allowing for integrability.

\section{Integrability tools}
\label{sec:int_tools}

It is well-known that matrix product states, a concept from quantum information theory, in combination with tools of integrability are efficient for the computation of planar one-point functions in the presence of defects in ${\cal N}=4$ SYM or ABJM theory, especially when the defects are associated with certain fields acquiring non-trivial VEVs. 
More precisely, the one-point function of a given bulk operator can be conveniently expressed in terms of the spin chain overlap between a Bethe eigenstate representing the operator and a certain matrix product state (MPS) encoding the classical background~\cite{deLeeuw:2015hxa,Buhl-Mortensen:2015gfd}. 
A general MPS takes the following form 
\begin{equation}
\label{eq:generalMPS}
    \langle\mbox{MPS}|=\tr \left(\sum_{I=1}^d\omega_I \langle I|\right)^{\!\otimes L},
\end{equation}
where $d$ is the dimension of the one-site Hilbert space on the spin chain, $|I\rangle$ are set of orthonormal basis vectors, and $\omega_I$ is a $k\times k$ matrix.
E.g.\ specialising to ${\cal N}=4$ SYM and to operators built from scalar
fields, i.e.\ the $SO(6)$ sector, the appropriate MPS for tree level computations takes the form
\begin{equation} \label{MPS}
\langle\mbox{MPS}|= 2^{\frac{L}{2}}\tr \left[\left( \phi_1^{cl}\;\langle 1| 
+\phi_2^{cl}\;\langle 2 | +\ldots + \phi_6^{cl} \;\langle 6 | \right)^{\otimes L}\right],
\end{equation}
where the trace is over the components of the classical background fields. 
The states $\langle 1|,\langle 2|, \ldots, \langle 6|$ form an orthonormal basis in the fundamental representation of an integrable $SO(6)$ spin chain. 
The parameter $k$ is identified with the size of the $\Fsu(2)$ representation in the definition of the rigid surface operator in \eq{rigidGW}.
Here, we have written the MPS in the basis of Hermitian fields, where the state $\langle I|$ is identified with the scalar $\phi_I$. 
An analogous expression exists for a basis of complex fields $\{|Z\rangle, |W\rangle,|Y\rangle,|\bar{Y}\rangle, |\bar{W}\rangle,|\bar{Z}\rangle\}$, 
where each complex field is a complex combination of the six real scalars.
Here, $|Z\rangle$ is the state associated with the scalar $Z$, and $\bar{Z}$ denotes the Hermitian conjugate of $Z$, and similarly for the other fields.
By convention $|Z\rangle$ is identified with the vacuum on a given site.
The vacuum state of the whole spin chain then is $|0\rangle\equiv|Z\rangle\otimes\cdots\otimes |Z\rangle$.

 A conformal single trace operator, ${\cal O}$, built from scalars can be identified with an eigenstate, $|\bf{u}\rangle$, of the integrable $SO(6)$ spin chain~\cite{Minahan:2002ve}.
 Its one-point function takes the form
\begin{equation} \label{onepointscaling}
    \langle {\cal O} \rangle=\frac{a_\CO}{r^{\Delta}},
\end{equation}
where $\Delta$ is the conformal dimension and $r$ is the distance to the defect.
At tree-level, $\Delta=L$, where $L$ is the number of fields entering the operator which is identified with the number of sites of the spin chain. 
Furthermore, the constant $a_\CO$ can be expressed as
\begin{equation}
    a_\CO={\cal C}_L \, \frac{\langle \mbox{MPS}|\bf{u}\rangle}{\,\,\,\,\langle {\bf u}| {\bf u}\rangle^{1/2}},
\end{equation}
where 
\begin{equation}
 {\cal C}_L=\left(\frac{2\pi^2}{\lambda}\right)^{L/2} \frac{1}{L^{1/2}},   
\end{equation}
is a field theoretical pre-factor which accounts for the difference between field theory normalisation of operators and the normalisation of spin chain eigenstates. 
Note that the one-point function is only defined up to a constant phase factor.

Certain classes of MPSs are integrable, allowing for their overlaps with Bethe eigenstates to be written in closed form. 
A straightforward way to check if a MPS is integrable is to relate it to a one- or two-site product state which obeys a so-called KT-relation~\cite{Piroli:2017sei,Pozsgay:2018dzs,Gombor:2021uxz,Gombor:2021hmj,Gombor:2024iix}. 
Here, $K$ refers to the K-matrix, i.e.\ the reflection amplitude of an excitation bouncing off the boundary of the spin chain.
$T$ refers to the monodromy matrix $T_A=\CL_{A,1}\ldots\CL_{A,L}$, which acts on all $L$ spin chain sites as well as on an auxiliary space denoted $A$.
The monodromy matrix satisfies the RTT-relation. 
Expressed in terms of the Lax operator $\CL$, it reads
\begin{equation}
\label{eq:RTT}
    R_{A,B}(u-v)\CL_{A,C}(u)\CL_{B,C}(v)=\CL_{B,C}(v)\CL_{A,C}(u)R_{A,B}(u-v)\,,
\end{equation}
where $R$ is the R-matrix of the spin chain, $B$ is an auxiliary space distinct from $A$, and $C$ labels an arbitrary spin chain site.
The KT-relation follows immediately from the requirement of invariance of the boundary state under the infinitely many odd charges (up to a spatial reflection across the boundary).
In the cases considered here it is sufficient to consider the action on one site only, and to only consider spatial reflections that act on $T_A(u)$ by reversing the sign of the spectral parameter $u$.\footnote{More generally, spatial reflections $\Pi$ may act non-trivially such that $\Pi T_A(u) \Pi\neq T_A(-u)$.
In this case the KT-relation is said to be crossed.
}
Such a KT-relation is said to be uncrossed.
For the MPS given in \eq{generalMPS}, the uncrossed KT-relation reads
\begin{equation}
\label{eq:KT}
K_{i,\ell}(u)\omega_I \CL_{\ell,j;I,J}(u)=\omega_I K_{\ell,j}(u)\CL_{i,\ell;I,J}(-u)\,,
\end{equation}
where repeated indices are summed over.
Here, $i,j,\ell=1,\ldots,D$ are indices over the $D$-dimensional auxiliary space, whereas $I,J=1,\ldots,d$ range over the one-site Hilbert space.
For a MPS, the components of $K$ and $\omega$ are $k\times k$ matrices, and matrix multiplication between them is understood.
Finding a non-trivial K-matrix is therefore equivalent to checking if the MPS is annihilated by all the parity-odd charges of the spin chain.

Recently, ref.~\cite{Gombor:2024iix} devised an efficient method to compute factorised overlaps involving MPSs provided there exists a non-trivial K-matrix that solves the KT-relation.
The resulting overlaps can be expressed as\footnote{
Note that in writing this expression, we make an implicit choice of conventions for the (nested) Bethe ansatz equations.
In particular, we adopt the conventions of ref.~\cite{Gombor:2024iix}.}
\begin{equation}
\label{eq:overlap_formula}
\frac{\langle\mbox{MPS}|\bf{u}\rangle}{\langle{\bf u}|{\bf u}\rangle^{1/2}}=\sum_{\gamma=1}^k \beta_\gamma\prod_{a=1}^{n_+}\prod_{j=1}^{n_a}\tilde{\CF}_\gamma^{(a)}(u_{a,j})\sqrt{\frac{\det G^+}{\det G^-}}\,,
\end{equation}
where the coefficients $\beta_\gamma$ are the eigenvalues of the overlap with the spin chain vacuum, $\langle\mbox{MPS}|0\rangle$. 
The functions $\tilde{\CF}^{(a)}_\gamma(u)$ have different definitions depending on the reflection algebra considered.
In the following sections, we will define and find explicit expressions for them, as needed.
The factor under the square-root is the superdeterminant of the Gaudin matrix.
$n_+$ labels the number of distinct types of Bethe roots $\{u^+_{a}\}=\{u_{a,1}, \ldots, u_{a,n_a}\}$, one for each $a=1,\ldots,n_+$, and $n_a$ labels the number of distinct roots in each set. 
The precise meaning of ``distinct'' will be clarified momentarily.
\emph{A priori}, each set of roots is associated with a node of the Dynkin diagram corresponding to the symmetry algebra $\Fg$ of the spin chain.
Naively, all roots can be taken to be different from one another, provided they solve the Bethe ansatz equations.
However, it was pointed out in ref.~\cite{Gombor:2020kgu} that the overlap $\langle\mbox{MPS}|{\bf u}\rangle=0$ unless Bethe roots satisfy one of two pairing conditions.\footnote{
In the following we assume that none of the roots are singular, i.e.\ $u_{a,j}\neq\iu/2$ for any $a$ and $j$.
Singular roots need to be treated separately, see ref.~\cite{Kristjansen:2021xno}.}

1. The first option requires that for every non-zero root $u_{a,j}$ in a given set $\{u^+_a\}$, there exists another root equal to $-u_{a,j}$.
If there is an odd number of roots, the remaining unpaired root must vanish.
If this holds for all sets $\{u^+_a\}$ associated with the nodes of the Dynkin diagram, then the reflection algebra has \emph{chiral pair structure}.
We identify $n_+=\text{rank}\, \Fg$ and $n_a=\lfloor \CN_a/2\rfloor$ where $\CN_a$ is the cardinality of the set $\{u^+_a\}$. All overlap formulas pertaining to domain wall
defects in ${\cal N}=4$ SYM are of this type~\cite{deLeeuw:2015hxa,deLeeuw:2016umh,DeLeeuw:2018cal,deLeeuw:2017cop,DeLeeuw:2019ohp}. The same is the case for 't Hooft lines~\cite{Kristjansen:2023ysz,Kristjansen:2024map}. 

2. For $\Fgl_N$ and $\Fo_{2N}$ only, there is an alternative pairing condition that exploits the $\mathds{Z}_2$ automorphism of these Lie algebras. 
This involution acts as a reflection on the corresponding Dynkin diagrams.
If nodes $a$ and $b$ of the Dynkin diagram are related by this involution, then the corresponding roots may obey $u_{a,j}=-u_{b,j}$.
Nodes that are left invariant by the $\mathds{Z}_2$ action must have all roots paired within each set.
This is referred to as \emph{achiral pair structure}. Overlap formulas for the domain wall version of ABJM theory are of this type~\cite{Kristjansen:2021abc,Gombor:2022aqj}.
Focussing on the $\Fo_{2p}$ case, we identify $n_+=p-1$, while $n_a=\CN_a$ for one of the two nodes that transform non-trivially under the involution and $n_a=\lfloor \CN_a/2\rfloor$ for the remaining $p-2$ nodes that are left invariant.

Symmetries determine which pair structures are allowed. 
They also encode how spatial reflections act on the monodromy matrix, which in turn determines whether the KT-relation is uncrossed as in \eq{KT}, or crossed.
The symmetry algebra of a spin chain is given by the Yangian algebra $Y(\Fg)$, and R-matrices are representations of $Y(\Fg)$.
Boundary conditions break the symmetry down to a subalgebra, forming a twisted Yangian $Y(\Fg,\Fh)$.
Here, $\Fh$ is a Lie subalgebra of $\Fg$, and the K-matrices are representations of $Y(\Fg,\Fh)$.
The allowed pair structure and the form of the KT-relation for each twisted Yangian were catalogued in ref.~\cite{Gombor:2024iix}.

\section{One-point functions for the ordinary defect \label{ordinary}}

We begin our integrability investigations by considering the $SO(6)$ and $SL(2)$ sectors of the ordinary Gukov-Witten defect.
As we will demonstrate, the $SO(6)$ sector admits a closed-form factorised expression for the overlap. 
The $SL(2)$ sector, however, turns out to be non-integrable.

\subsection{$SO(6)$ sector}

In the Abelian case the MPS is fully reducible.
This implies that the trace in eq.~(\ref{MPS}) reduces to a sum over powers of diagonal elements of the matrix in eq.~(\ref{ZAbelian}). 
Thus the relevant MPS for single trace operators built from scalars, i.e.\ operators in the $SO(6)$ sector, can be expressed as 
\begin{equation}
\langle \mbox{MPS}|= \sum_{m=1}^M N_m
\left(\langle Z|\, \frac{\beta_m+\iu\gamma_m}{ e^{\iu\psi}}+ \langle \bar{Z}|\, \frac{\beta_m-\iu\gamma_m}{e^{-\iu\psi}}\right)^{\!\otimes L}.
\end{equation}
Here we define the complex fields by $Z=\Phi=\frac{1}{\sqrt{2}}(\phi_1+\iu \phi_2)$, and then the corresponding basis is given by
\begin{equation}   
\langle Z|= \frac{1}{\sqrt{2}}(\langle 1|-\iu \langle 2|)\,,
\end{equation}
where $\langle I|=(|I\rangle)^\dagger$ for $I=1,\ldots,6$.
For convenience we define $\psi_m=\psi-\chi_m$, where $e^{\iu\chi_m}=(\beta_m+\iu\gamma_m)/\sqrt{\beta_m^2+\gamma_m^2}$, and rewrite the MPS as
\begin{equation}
\langle \mbox{MPS}|= \sum_{m=1}^M N_m
(\beta_m^2+\gamma_m^2)^{L/2} 
\left(\langle Z|\, \, e^{-\iu\psi_m}+ \langle \bar{Z}|\, e^{\iu\psi_m}\right)^{\!\otimes L}.
\end{equation}
In order for the overlap to be non-vanishing, the Bethe eigenstate has to be built entirely from $|Z\rangle$'s and $|\bar{Z}\rangle$'s which implies that the following selection rule has to be fulfilled
\begin{equation}
\label{eq:select_ruleSO6}
\CN_1=2\CN_2=2\CN_3,
\end{equation}
where $\CN_a$ is the total number of Bethe roots associated with node $a$ of the Dynkin diagram of $SO(6)$. 
Here, $a=1$ denotes the middle node and $a=2,3$ denote the two edge nodes.
In particular we can replace the MPS above with the even simpler
one
\begin{equation}
\begin{split}
\langle \mbox{MPS}|= 2^\frac{L}{2}\sum_{m=1}^M& N_m \,
(\beta_m^2+\gamma_m^2)^{L/2}\times\\
&
(\langle 1|\cos\psi_m-\langle 2|\sin\psi_m)^{\otimes L}.
\end{split}
\end{equation}
Because our MPS just involves a sum over ordinary boundary states without matrix structure, we can consider each term in the sum over $m$ separately.
Each contribution from a given site can be identified with a vector in the 6d R-symmetry space, whose stabiliser is $SO(5)$.
Therefore, the appropriate reflection algebra is the twisted Yangian $Y(\Fo_6,\Fo_5)$, leading to an uncrossed KT-relation and chiral pair structure.
The overlap between a closely related MPS and the Bethe eigenstates of the integrable $SO(6)$ spin chain was determined in~\cite{DeLeeuw:2019ohp}, where it was also
shown that the root configuration must have chiral pair structure.
The quantity of relevance for the one-point function can be read off to
be 
\begin{equation}
\label{eq:ordinarySO6}
\begin{split}
 \frac{\langle \mbox{MPS}|\bf{u}\rangle}{\,\,\,\,\langle {\bf u} |{\bf u}\rangle^{1/2}}&= \sum_{m=1}^M N_m e^{-\iu (L-{\cal N}_1)\psi_m}
(\beta_m^2+\gamma_m^2)^{L/2} \, \times  \\
 &\sqrt{\frac{Q_1(0)Q_1(\iu/2)}{Q_2(0)Q_2(\iu/2)Q_3(0)Q_3(\iu/2)}\frac{\det G^+}{\det G^-}},
\end{split}
\end{equation}
where the $Q_a$ is the Baxter polynomial associated with the $a^{\text{th}}$ node, $Q_a(u)=\prod_{j=1}^{n_a}(u^2-u^2_{a,j})$ with zero roots omitted.
The Bethe roots $\{u_{a}^+\}$ are solutions to the nested Bethe ansatz equations, see e.g.\ ref.~\cite{deLeeuw:2016ofj} for an explicit expression in the conventions used here.
The superdeterminant of the Gaudin matrix is a function of the Bethe roots, and it can be found e.g.\ in ref.~\cite{Kristjansen:2021xno}.

Note that the $SU(2)$ sector is reached by taking $\CN_2=\CN_3=0$.
By the selection rule \eq{select_ruleSO6}, we must have that $\CN_1=0$, and hence we see that in this sector only the vacuum state has a non-vanishing overlap.

\subsection{$SL(2)$ sector}
\label{sec:ordinarySL2}

Unlike the $SO(6)$ sector, the $SL(2)$ sector is closed beyond one-loop order.
It consists of single trace operators built out of one species of complex field acted on by light-cone covariant derivatives. 
Due to the abelian nature of the VEVs the covariant derivatives reduce to ordinary derivatives.
A convenient choice of light-cone direction corresponds to $D=\partial_x+\partial_t$. 
First, we notice that for each diagonal element of $Z=\Phi$ we have
\begin{equation}
\partial_x^\ell\, Z \,\propto \, \partial_x^\ell \frac{e^{-\iu\psi}}{r}=(-1)^\ell \ell! \, \frac{e^{-\iu(\ell+1)\psi}}{r^{\ell+1}}\,.
\end{equation}
The $\psi$-dependent phase only gives rise to an overall pre-factor in the one-point functions and the remaining part of the derivative coincides with the derivative that appears in the $SL(2)$ sector of the D3-D5 domain wall version of ${\cal N}=4$ SYM.
For that system the overlaps are known to vanish~\cite{Zarembo}.

Alternatively, one can consider a different $SL(2)$ sub-sector by associating the VEVs with complex fields in another way
\begin{equation}\label{B2 VEVs Z}
Z=\frac{1}{\sqrt{2}}(\phi_1+\iu \phi_3),
\end{equation}
where $\phi_3$ has vanishing VEV.
To compute the one-point functions in this sub-sector one should consider
the overlap of Bethe eigenstates with the boundary state $\langle B|=\sum_{m=1}^M N_m (\langle\CB_m|)^{\otimes L}$, where
\begin{equation}
\label{eq:B2}
\langle \CB_m |=\sum_{\ell=0}^\infty \langle \ell| (-1)^\ell \frac{\beta_m\cos((\ell+1)\psi)+\gamma_m\sin((\ell+1)\psi)}{ r^{\ell}}\,,
\end{equation}
and $|\ell\rangle \leftrightarrow \frac{1}{l!} \partial_x^{\ell} Z$.
This is analogous to the computation carried out for one-point functions in the background of the 't Hooft line in refs.~\cite{Kristjansen:2023ysz,Gombor:2023bez}. 
In the latter case, the factor multiplying $\langle \ell|$ takes the form of the $\ell^{\text{th}}$ Legendre polynomial. 
The recursion relation of the Legendre polynomials was instrumental in proving the integrability of the boundary state there.
In the present case of the Gukov-Witten defect, however, the pre-factor does not allow for a similar proof.
What is more, one can show that the first of the odd conserved charges of the $SL(2)$ spin chain does not annihilate the boundary state $| B\rangle$.
Hence the boundary state is not integrable, and we cannot find a closed overlap formula in the present case.

\section{One-point functions for the rigid defect \label{rigid}}

The rigid defect has the common feature with the thoroughly studied probe D3-D5 brane set-up that the classical fields are expressed in terms of $SU(2)$ generators. 
One may therefore hope that some of the results derived for the D3-D5 system can be applied to the present one. 
As we show in this section, this is only the case for the $SU(2)$ sector, whereas the $SO(6)$ and $SL(2)$ sectors differ substantially. 
We find that the $SO(6)$ and $SL(2)$ sectors are integrable only if the dimension of the $\Fsu(2)$ representation $k=2$.
We present closed-form expressions for the overlaps.

\subsection{$SU(2)$ sector}

Suppose we identify the complex fields $Z$ and $W$ as follows 
\begin{equation}\label{B2 VEVs Z, W}
Z=\frac{1}{\sqrt{2}}(\phi_1+\iu\phi_3), \hspace{0.5cm} W=\frac{1}{\sqrt{2}}({\phi_2+\iu\phi_4}).
\end{equation}
The resulting $SU(2)$-subsector has non-vanishing overlaps, which were determined in closed form for any dimension, $k$, of the $SU(2)$ representation in~\cite{DeLeeuw:2018cal}. 
The result for even $L$ and an even number of excitations is
    \begin{equation}\label{MPSk-u}
\frac{|\langle \mbox{MPS}|\bf{u}\rangle|}{\,\,\,\,\langle {\bf u} |{\bf u}\rangle^{1/2}}=
 \mathbb{S}_{k}\,Q\left(\frac{\iu k}{2}\right)\sqrt{Q(\iu/2)Q(0)\,\frac{\det G^+}{\det G^-}}\,,
\end{equation}
where 
\begin{equation}\label{Tk}
 \mathbb{S}_k=\sum_{q=-\frac{k-1}{2}}^{\frac{k-1}{2}}\frac{q^L}{Q\left(\frac{2q+1}{2}\,\iu\right)Q\left(\frac{2q-1}{2}\,\iu\right)}\,,
\end{equation}
and zero otherwise.
An extension of the corresponding one-point function formula to all loop
orders was presented in~\cite{Gombor:2020kgu,Komatsu:2020sup}. 
We note that in the present case the general formula (\ref{onepointscaling}) gets an extra factor of $(\log(\frac{r}{r_0}))^{-L}$. 

\subsection{$SO(6)$ sector}

Clearly, the result for the $SO(6)$ sector cannot carry over from the D3-D5 system. 
This is due to the fact that in the present case only two out of the six scalar fields acquire a VEV, whereas for the D3-D5 system three scalars had a non-trivial background.

A quick way to test if a given MPS is integrable is to check if it is annihilated by the first in the chain of odd conserved charges of the spin chain, $Q_3$. 
The action of $Q_3$ of the $SO(6)$ spin chain on a general MPS was spelled out in detail in~\cite{deLeeuw:2016ofj}. 
When applied to the MPS of relevance for the $SO(6)$ sector of the present set-up, one encounters the complication that the absence of one of the three generators of $\su(2)$ prevents us from making use of the Casimir relation of the algebra. 
Instead, the action of the charge $Q_3$ involves the anti-commutator $\{t_i,t_j\}$ with $i,j=1,2,3$. 
As a result, $Q_3$ only annihilates the MPS for the two-dimensional representation with $k=2$, for which the anti-commutator is proportional to the identity matrix. 
For this representation one can in addition show that the MPS is annihilated by the entire tower of odd charges by appropriately adjusting the argument given in~\cite{DeLeeuw:2018cal}. 
In the following we shall explicitly demonstrate integrability by solving the KT-relation, and determine the overlaps in closed form.

Since the VEVs in eq.~\eqref{ZAbelian} preserve an $SO(2)\times SO(4)$ subgroup of the $SO(6)$ R-symmetry, we can consider a spin chain with reflection algebra given by the twisted Yangian $Y(\Fo(6),\Fo(2)\oplus\Fo(4))$.\footnote{Equivalently, one can work with $Y(\Fgl(4),\Fgl(2)\oplus\Fgl(2))$ which encodes the same symmetries after quotienting by an overall $\Fu(1)$.}
For this reflection algebra, the KT-relation is uncrossed as in \eq{KT} and the Bethe roots have achiral pair structure.
Note that this implies the selection rule $\CN_2=\CN_3$ on the number of Bethe roots.
The auxiliary space is chosen to have the same dimension as the one-site Hilbert space, $d=D=6$, and we will use indices $I,J$ to denote either.
The Lax operator entering \eq{KT} may be taken to be $\CL(u)=R(u+\xi)$, where the R-matrix of the $\Fso(6)$ spin chain is given by
\begin{equation}
    R(u)=\mathds{1}+\frac{1}{u}\bm{P}-\frac{1}{u-2}\bm{K}\,,
\end{equation}
and the permutation and trace operators are defined as 
\begin{align}
    \bm{P}&=\sum_{I,J=1}^6 e_{I,J}\otimes e_{J,I}\,,& \bm{K}&=\sum_{I,J=1}^6 e_{I,J}\otimes e_{7-I,7-J}\,,
\end{align}
respectively.
Here, we work in the basis $\{Z,W,Y,\bar{Y},\bar{W},\bar{Z}\}$, where we define the complex fields as in eq. (\ref{B2 VEVs Z, W}) together with $Y=\frac{1}{\sqrt{2}}(\phi_5+\iu\phi_6)$. The \emph{a priori} free parameter $\xi$ arises due to the invariance of the RTT-relation \eq{RTT} under a constant shift of $u,v$ by any $\xi$. 
The KT-relation \eq{KT}, however, does not inherit this shift-invariance, and the parameter $\xi$ must be fixed appropriately.
The overlap of the KT-relation \eq{KT} for components $i= 6$ and $j=1$ with the spin chain vacuum $|0\rangle$ leads to a symmetry relation on the vacuum eigenvalues of the monodromy matrix 
\begin{equation}
    \lambda_I(u)=\lambda_{7-I}(u)\,,
\end{equation}
where $T_{I,I}(u)|0\rangle = \lambda_I |0\rangle$ and $I=1,\ldots,6$.
The eigenvalues $\lambda_I$ depend on the shift $\xi$.
The symmetry relation is only obeyed if $\xi=-1$, thus fixing the ambiguity.

The KT-relation \eq{KT} with $\CL(u)=R(u-1)$ can now be straightforwardly solved for $K$. 
Substituting the VEVs of the rigid defect given in \eq{rigidGW} into the general form of the MPS in eq.~\eqref{MPS}, i.e.\ $\omega_I=(\phi_1,\phi_2,0,0,\phi_2,\phi_1)$, we find the following solution to the KT-relation
\begin{equation}
K(u)=    \begin{pmatrix}
\mathds{1}_2 & \iu\sigma_3 & 0 & 0 & \iu\sigma_3 & U\\
-\iu \sigma_3 & \mathds{1}_2 & 0 & 0 & U & -\iu\sigma_3\\
0 & 0 & -U+\mathds{1}_2 & 0 & 0 & 0\\
0 & 0 & 0 & -U+\mathds{1}_2 & 0 & 0\\
-\iu\sigma_3 & U & 0 & 0 & \mathds{1}_2 & -\iu\sigma_3\\
U & \iu\sigma_3 & 0 & 0 & \iu\sigma_3 & \mathds{1}_2
\end{pmatrix},
\end{equation}
where $U \equiv (2u-1)\mathds{1}_2$.
In order to apply the overlap formula \eq{overlap_formula}, one is instructed to define a set of nested K-matrices that reflect the nested nature of the Bethe ansatz~\cite{Gombor:2020kgu}.
For convenience, we relabel the components of the K-matrix in the auxiliary space such that $I,J=-3,-2,-1,1,2,3$ in the following.\footnote{Concretely, we relabel $\{1,2,3,4,5,6\}\to\{-3,-2,-1,1,2,3\}$, such that e.g.\ $K_{1,1}$ in the notation of section \ref{sec:int_tools} maps to $K_{-3,-3}$ here.}
The K-matrix at the second level of nesting is then defined as
\begin{equation}
    K^{(2)}_{I,J}(u)=K_{I,J}(u)-K_{I,-3}(u)\left(K_{3,-3}(u)\right)^{-1}K_{3,J}(u).
\end{equation}
With these K-matrices, we further define
\begin{equation}
\begin{split}
    \bm{G}^{(1)}&=K_{3,-3}(u)\,, \qquad\qquad\qquad\;\; \bm{G}^{(2)}=K^{(2)}_{2,-2}(u)\,,\\
    \bm{G}^{(3)}&=K^{(2)}_{1,1}(u)-K^{(2)}_{1,-2}(u)\left(K^{(2)}_{2,-2}(u)\right)^{-1}K^{(2)}_{2,1}(u)\,,
\end{split}
\end{equation}
and
\begin{equation}
\begin{split}
\bm{F}^{(1)}&=\left(\bm{G}^{(1)}\right)^{-1}\bm{G}^{(2)}=\frac{4u(u-1)}{2u-1}\mathds{1}_2\,,\\
\bm{F}^{(2)}&=\left(\bm{G}^{(2)}\right)^{-1}\bm{G}^{(3)}=\frac{1-2u}{2u}\mathds{1}_2\,.
\end{split}
\end{equation}
Denoting the (degenerate) eigenvalues of $\bm{F}^{(a)}(u)$ by $\CF^{(a)}(u)$ for $a=1,2$, the functions $\tilde{\CF}^{(1)}(u)$ entering the overlap formula \eq{overlap_formula} are 
\begin{align}
\tilde{\CF}^{(1)}(u)&=\CF^{(1)}(\iu u + 1/2)\sqrt{\frac{u^2}{u^2+1/4}}=\sqrt{\frac{u^2+1/4}{u^2}},\\
\tilde{\CF}^{(2)}(u)&=\CF^{(2)}(\iu u)\frac{u}{u+\iu/2}=-1.
\end{align}
The resulting overlap then is 
\begin{equation}
 \frac{\langle \mbox{MPS}|\bf{u}\rangle}{\langle {\bf u} |{\bf u}\rangle^{1/2}}=
 \frac{1}{2^{L-1}}   \sqrt{\frac{Q_1(\iu/2)}{Q_1(0)} \, \frac{\det G^+}{\det G^-}}
\end{equation}
for even $L$. 
For odd $L$ the overlap vanishes because the eigenvalues $\beta_\gamma$ of $\langle \mbox{MPS}|0\rangle$ are equal in magnitude but have opposite sign.
The Gaudin determinants entering this expression assume achiral pair structure. 
Explicit expressions for them in our conventions can be found in ref.~\cite{Kristjansen:2021abc}.

Note that even though only $Q$-functions for the central node appear in the overlap, there is an implicit dependence on the Bethe roots for the edge nodes through the Gaudin factor.
Note that our formula closely resembles the one found numerically by refs.~\cite{Jiang:2019xdz, Jiang:2019zig} for the three-point function of two determinant and one single-trace operator in $\CN=4$ SYM theory without defects.
The two set-ups share the same symmetries, and their MPSs are closely related. 
We here provide an analytic derivation of the overlap.
Moreover, our formula is analogous to the one found for the ABJM domain wall with an alternating $\su(4)$ spin chain, which also features achiral pair structure and dependence on one of the two distinct sets of Bethe roots only~\cite{Kristjansen:2021abc}.
Achiral pair structure also appears in three-point functions in ABJM theory without defects~\cite{Yang:2021hrl}.

\subsection{$SL(2)$ sector}

Compared with the discussion of section~\ref{sec:ordinarySL2}, the $SL(2)$ sector for the rigid defect appears more involved due to the appearance of the logarithm in \eq{rigidGW}.
Successive light-cone partial derivatives of these VEVs give rise to ever more complicated functions of $r$.
However, the leading singularity arises from the partial derivatives acting on $1/r$ only, whereas derivatives acting on $1/\log r$ lead to terms with a milder singularity.
The definition of the defect \eq{rigidGW} is agnostic to any potential sub-leading singularities. 
We may therefore restrict our discussion of the $SL(2)$ sector to the leading singularity only, which is proportional to $1/\log r$.
Passing from partial derivatives $\partial_x$ to covariant derivatives $D_x$ leads to contributions involving the commutator $[A_x,Z]$.
However, since $A_x\propto 1/\log r$, we may consistently drop these contributions, i.e.\ the commutator terms are sub-leading. 
At the leading singularity, with the identification as in eq.\ (\ref{B2 VEVs Z}), one finds the boundary state $\langle \text{MPS} |=\text{tr}(\langle \mathcal{B} |^{\otimes L})$ where $\langle \mathcal{B} |$ has the compact expression
\begin{equation}\label{eq: SL2 matrix state}
    \langle \CB | = \sum_{\ell=0}^{\infty} \langle \ell |(-1)^{\ell} \frac{t_1\cos((\ell+1)\psi)+t_2\sin((\ell+1)\psi)}{r^{\ell}
    }.
\end{equation}
The symmetry preserved in the $SL(2)$ sector can be identified with the twisted Yangian $Y(\mathfrak{gl}_2, \mathfrak{gl}_1\oplus\mathfrak{gl}_1)$ after quotienting by an overall $\Fu(1)$. 
We may now proceed with the method of ref.~\cite{Gombor:2024iix}.
Following ref.~\cite{Gombor:2023bez}, we take the Lax operator entering the KT-relation \eq{KT} to be
\begin{equation}
    \CL(u)=\mathds{1}+\frac{1}{u}\sum_{i,j=-1,1}e_{i,j}\otimes E_{j,i}\,,
\end{equation}
where $i,j$ are indices over the auxiliary space of dimension $D=2$.
We fixed the invariance under rescalings of the spectral parameter $u$ by adopting the conventions of ref.~\cite{Kristjansen:2023ysz} for the $SL(2)$ Bethe ansatz equations. 
The $2\times 2$ matrix $e_{i,j}$ acting on the auxiliary space is defined as $(e_{i,j})_{k,\ell}=\delta_{i,k}\delta_{j,\ell}$, while the $\mathfrak{sl}_2$ representation $E_{i,j}$ is chosen as
\begin{equation}
\begin{split}
    E_{-1,-1}|\ell\rangle&=-(\ell+1/2)\,|\ell\rangle\,,\\
    E_{-1,1}|\ell\rangle&=-\iu\ell\,|\ell-1\rangle \,,\\
    E_{1,-1}|\ell\rangle&=-\iu(\ell+1)\,|\ell+1\rangle\,,\\
    E_{1,1}|\ell\rangle&=(\ell+1/2))\,|\ell\rangle\,.
\end{split}
\end{equation}
With this Lax operator and taking the size of the $\su(2)$ representation $k=2$, we find the following solution to the KT-relation 
\begin{equation}
    K(u)= \begin{pmatrix}
\sin\psi\sigma_3 - 2\iu u \cos\psi & -2u/r\\
-2ru &  \sin\psi\sigma_3 + 2\iu u \cos\psi
\end{pmatrix}\!,
\end{equation}
where each entry is a $2\times2$ matrix.
For $k\geq3$, there are no non-trivial solutions to the KT-relation, which we checked up to $k=8$.
This is consistent with the observation that $Q_3$ does not annihilate the boundary state unless $k=2$.
Denoting the entries of the K-matrix $K_{i,j}$ with $i,j=-1,1$, we define
\begin{equation}
    \begin{split}
        \bm{G}^{(1)}&=K_{1,-1}\,,\\
        \bm{G}^{(2)}&=K_{-1,1}-K_{-1,-1}(K_{1,-1})^{-1}K_{1,1}\,,
    \end{split}
\end{equation}
such that
\begin{equation}
            \bm{F}^{(1)}(u)=\left(\bm{G}^{(1)}\right)^{-1}\bm{G}^{(2)}=\frac{u^2-1/4}{u^2}\frac{\sin^2\psi}{r^2}\mathds{1}_2\,.
\end{equation}
Again, denoting the (degenerate) eigenvalues of $\bm{F}^{(1)}(u)$ as $\CF^{(1)}(u)$, we find that the function entering the overlap formula is 
\begin{equation}
    \tilde{\CF}^{(1)}=\CF^{(1)}(\iu u)\sqrt{\frac{u^2}{u^2+1/4}}=\sqrt{\frac{u^2+1/4}{u^2}}\frac{\sin^2\psi}{r^2}\,.
\end{equation}
The corresponding overlap for even $L$ is therefore 
\begin{equation}\label{eq: SL2 rigid overlap}
 \frac{\langle \mbox{MPS}|\bf{u}\rangle}{\,\,\,\,\langle {\bf u} |{\bf u}\rangle^{1/2}}= \frac{\sin^{\CN} \psi}{r^{\CN}
 }\frac{1}{2^{L-1}}\sqrt{\frac{Q(\iu/2)}{Q(0)}\frac{\det G^+}{\det G^-}},
\end{equation}
where $\CN$ denotes the cardinality of the set of Bethe roots. 
For odd $L$ the overlap vanishes for the same reason as in the $SO(6)$ sector.

This formula is similar to the one found numerically by refs.~\cite{Jiang:2019xdz, Jiang:2019zig} for the $SL(2)$ sector of three-point functions with two determinant operators in $\CN=4$ SYM theory.

\section{Outlook\label{Outlook}}

In this work we investigated integrability properties of Gukov-Witten defects. 
We showed that the ordinary Gukov-Witten defect is generically non-integrable, except for the $SO(6)$ sector where the overlap can be written in closed form.
For the rigid Gukov-Witten defect, we showed that $SO(6)$ and $SL(2)$ sectors are both integrable only if the dimension of the $\su(2)$ representation that enters the definition of the operator is $k=2$.
Only the $SU(2)$ sector was found to be integrable also for $k>2$.
We reported closed-form factorised expressions for all sectors.

It would be interesting to investigate whether the integrability that we have identified in certain corners of the space of Gukov-Witten defects allows for extension in any direction. 
The most promising case seems to be the extension of the rigid surface case with $k=2$ to further sub-sectors and to higher loop orders. 
For the D3-D5 domain wall in ${\cal N}=4$ SYM and the similar D2-D4 domain wall in ABJM theory a simple argument based on covariance of overlaps under fermionc dualities allowed an educated proposal for the full tree-level overlap formula~\cite{Kristjansen:2020vbe,Kristjansen:2021abc}. 
Furthermore, for the former case one could argue for an extension to an all-loop asymptotic overlap formula by a combination of symmetry considerations and integrability bootstrap~\cite{Gombor:2020kgu,Gombor:2020auk,Komatsu:2020sup}. 
The same strategy could be applied to the 't Hooft loop in ${\cal N}=4$ SYM~\cite{Gombor:2024api}.
For $\CN=4$ SYM theory without defects, refs.~\cite{Jiang:2019xdz, Jiang:2019zig} obtained a closed-form expression for the three-point function of two determinant operators and one single-trace operator that is exact in the 't Hooft coupling.
As noted in the previous section, one-point functions of single-trace operators with a rigid Gukov-Witten defect appear to be closely related to that set-up.
This suggests that it may be possible to find an all-loop formula for the one-point function with $k=2$.

Recently, ref.~\cite{Choi:2024ktc} set up the perturbative program for higher loop computations in the background of the ordinary Gukov-Witten defect.
The one for the rigid case should follow from the one of the D3-D5 probe brane case~\cite{Buhl-Mortensen:2016pxs,Buhl-Mortensen:2016jqo}.
It would be interesting to investigate if SUSY localisation in combination with perturbative results for two-point functions could be used as input to the boundary conformal bootstrap program along the lines of~\cite{deLeeuw:2017dkd,Baerman:2024tql}.

Ref.~\cite{deLeeuw:2024qki} pointed out another 1/2-BPS surface defect in $\CN=4$ that has received considerably less attention.
The defect preserves chiral 2d $\CN=(0,8)$ SUSY and is the field theory dual of the holographic set-up of ref.~\cite{Harvey:2008zz}. 
This system's integrability properties likewise deserves further investigation.

\vspace{0.3cm}

\section*{Acknowledgments}
We would like to thank Jonah Baerman, Shota Komatsu and Konstantin Zarembo for useful discussions. 
We are particularly grateful to Tamas Gombor for sharing his unique insights with us, and to Jonah Baerman, Tamas Gombor and Kostantin Zarembo for comments on the draft.
A.C.\ and C.K.\ were supported by DFF-FNU through grant number 1026-00103B. 
C.S was supported by GEP through grant number JY202211.

\bibliographystyle{elsarticle-num} 

\bibliography{GW}

\end{document}